\documentclass[12pt]{article}

\usepackage{bm}% bold math
\usepackage{amssymb}
\usepackage{enumerate}

\usepackage{graphicx} 
\begin{document}

%\maketitle
\title{\bf The muon (g - 2) anomaly and a new light vector boson} 

\author{ N.V.~Krasnikov$^{1,2}$
\\
\\
$^{1}$ INR RAS, 117312 Moscow, Russia
\\
$^{2}$ Joint Institute for Nuclear Research, 141980 Dubna, Russia} 

%%$  $DOI:10.3204/DESY-PROC-2016-04/krasnikov-nikolai                   

%\preprint{APS/123-QED}

%\title{Manuscript Title:\\with Forced Linebreak}% Force line breaks with \\
%\title{Search for light dark vector boson. NA64 experiment } 
%and constraints on new physics}
%\title{The  MiniBooNE anomaly and heavy neutrino decay}

\date{        } % It is always \today, today,
%\date{June 17, 2009}% It is always \today, today,
\maketitle

\begin{abstract}
This mini review aims to outline the main experimental and theoretical results related with the search for a new light vector boson  
proposed as a solution of the muon $g-2$ anomaly. Additionally,  I consider a model with infinite number of 
light vector bosons which can explain the anomaly, while still remaining consistent with   current experimental bounds. 
\end{abstract}

%\pacs{14.80.-j, 12.60.-i, 13.20.Cz, 13.35.Hb}

\newpage

\section{Introduction}
The search for new physics beyond the SM can be divided into two categories. The first one is the search for new heavy particles and 
interactions at high energies, 
the so-called ``energy frontier'' research. Typical examples are LEP, Tevatron and LHC. The second type of research is the search for relatively 
light with masses $m \leq O(1)~GeV$ new hypothetical particles. In this case an experiment needs to cross 
the ``intensity frontier''. The most famous example of 
light hypothetical particle is the axion \cite{axion}, invented for the solution of strong CP-problem. Also there are models predicting the existence of 
light scalar, spin $1/2$ and vector particles. In particular, models with light vector bosons \cite{REVVEC} (vector portal) become 
 rather popular now. Light vector boson  $Z'$   can  mediate between our world and dark sector \cite{REVVEC}. Also 
 $Z'$  can explain \cite{vecmuon1} - \cite{vecmuon6}  the muon $(g-2)$ anomaly \cite{g-2}. Recent claim \cite{17mev} 
of the discovery of  $~17~MeV$ vector particle observed 
as a peak in $e^+e^-$ invariant mass distribution in nuclear transitions makes the question of possible $Z'$  existence  
extremely interesting and important and enhance motivation for the experimental searches at 
low energy intensity frontier. 
In this mini review I  outline the main experimental and theoretical results related with the search of light $Z'$. Additionally,  
I consider a model with infinite number of 
light vector bosons which can explain the anomaly, while still remaining consistent with   current experimental bounds.

\section{Muon  $g-2$ anomaly and the light $Z'$.}

Recent precise measurement of the  anomalous magnetic 
moment of the positive muon $a _{\mu}= (g-2)/2$ from Brookhaven AGS experiment 821 \cite{g-2} gives result which is about
$3.6 \sigma$ higher \cite{g-2.th, dorokhov} than the 
Standard Model prediction 
\begin{equation}
a^{exp}_{\mu} -a^{SM}_{\mu} = 288(80)\times 10^{-11}  \,.
\end{equation}
This result may signal the existence of new physics beyond the 
Standard Model. New light  (with a mass $m_{Z'} \leq O(1)~GeV$)
vector boson (dark photon) which couples very weakly with muon with $\alpha_{Z'} \sim O(10^{-8})$ 
can explain $(g-2)$ anomaly \cite{vecmuon1} - \cite{vecmuon6}.  Vectorlike interaction of $Z'$ boson with muon  
\begin{equation}
L_{Z'} = g^`\bar{\mu}\gamma^{\mu}\mu Z'_{\mu}\,
\end{equation}
leads to  additional contribution to  muon 
anomalous magnetic moment \cite{MuonMoment}     
\begin{equation}
\delta a  = \frac{\alpha^`}{2\pi} F(\frac{m_{Z'}}{m_{\mu}}) \,,
\end{equation}
where
\begin{equation}
F(x) = \int^1_0 dz \frac{[2z(1-z)^2]}{[(1-z)^2 + x^2z]} \,
\end{equation}
and $\alpha^` = \frac{(g^`)^2}{4\pi}$. Equation (4) allows to determine the coupling constant $\alpha^`$ which 
explains the value (1) of muon anomaly. For $ m_{Z'} \ll m_{\mu}$ one can  find that
\begin{equation}
\alpha^` = (1.8 \pm 0.5) \times 10^{-8}\\.
\end{equation}
For another limiting case  $ m_{Z'} \gg m_{\mu}$ the $\alpha^`$ is 
\begin{equation}
\alpha^` = (2.7 \pm 0.7) \times 10^{-8} \times \frac{m^2_{Z'}}{m^2_{\mu}}   \\.
\end{equation}

But the postulation of the interaction (2)   is not the end of the story. 
The main question: what about the interaction of the  $Z'$  boson with other quarks and leptons? 
The  $Z'$ interaction with the SM fermions $\psi_{k}$ $(\psi_k = e, \nu_e, u, d, ...)$ 
has the form
\begin{equation}
L_{Z'} = g^`Z'_{\mu}J_{Z'}^{\mu}\\,
\end{equation}
\begin{equation}
J_{Z'}^{\mu} = \sum_{k}[q_{Lk}\bar{\psi}_{Lk}\gamma^{\mu}\psi_{Lk} + q_{Rk}\bar{\psi}_{Rk}\gamma^{\mu}\psi_{Rk}] \\,
\end{equation}
where  $\psi_{Lk,Rk} = \frac{1}{2}(1 \mp \gamma_5)\psi$ and 
$q_{Lk}, q_{Rk}$ are the $Z'$ charges of the $\psi_{Lk}, \psi_{Rk}$ fermions.
The $Z'$ can interact with   new hypothetical particles beyond the SM, for instance, with  dark matter fermions $\chi$
\begin{equation}
L_{Z',\chi} = g_D Z'_{\mu}\bar{\chi}\gamma^{\mu}\chi \\.
\end{equation}
There are several models of the current $J_{Z'}^{\mu}$. In a model \cite{X1,X2} $Z'$ interacts with
photon $A_{\mu}$  due to 
kinetic mixing term\footnote{Here $F_{\mu\nu} = \partial_{\mu}A_{\nu} - \partial_{\nu}A_{\mu}$ and 
$Z'_{\mu\nu} = \partial_{\mu}Z'_{\nu} - \partial_{\nu}Z'_{\mu}$.} 
\begin{equation}
L_{mix} = \frac{\epsilon}{2}  F^{\mu \nu} Z'_{\mu \nu} \\.
\end{equation}
 As a result of the mixing (10) the field $Z'$ interacts with the SM electromagnetic field 
$J^{\mu}_{EM} = \frac{2}{3}\bar{u}\gamma^{\mu}u - \frac{1}{3}\bar{d}\gamma^{\mu}d - \bar{e}\gamma^{\mu}e + ...$
and the  coupling constant  
$g^` = \epsilon e$ ($\alpha = \frac{e^2}{4\pi} = \frac{1}{137}$).                              .
Other interesting scenario is the model
\cite{LEE} where $Z'$ 
(the dark leptonic gauge boson)  
interacts with the SM leptonic current, namely 
\begin{equation}
L_{Z'} = g'[\bar{e}\gamma^{\nu}e + \bar{\nu}_{eL} \gamma^{\nu}\nu_{eL} + \bar{\mu}\gamma^{\nu}\mu +  
\bar{\nu}_{\mu L} \gamma^{\nu}\nu_{\mu L} \nonumber \\
+ \bar{\tau}\gamma^{\nu}\tau +  \bar{\nu}_{\tau L} \gamma^{\nu}\nu_{\tau L}]Z'_\nu  \,. 
\end{equation}

In Refs. \cite{vecmuon1} - \cite{vecmuon3} for an explanation of $g - 2$ muon anomaly a model where 
$Z'$ interacts predominantly with the second and third generations through the  
$L_{\mu} - L_{\tau}$ current 
\begin{equation}
L_{Z'} = g'[\bar{\mu}\gamma^{\nu}\mu +  \bar{\nu}_{\mu L} \gamma^{\nu}\nu_{\mu L}
- \bar{\tau}\gamma^{\nu}\tau  -  \bar{\nu}_{\tau L} \gamma^{\nu}\nu_{\tau L}]Z'_{\nu}
\end{equation}
has been proposed.
The interaction (12) is $\gamma_5$-anomaly free, 
it commutes with the SM gauge group  and moreover it  escapes (see next section) from the most restrictive current 
experimental bounds because the interaction (12) does not contain quarks and first generation leptons $\nu_e$, $e$.

In Ref.\cite{NKSG} a model where $Z^`$ couples with a right-handed current of the first 
and second generation  SM fermions including the right-handed neutrinos has been suggested. The model is 
able to explain the muon $(g-2)$ anomaly due to existence of light scalar and can be tested in future experiments.

%%%%%%%%%%%%%%%%%%%%%%%REMOVE

\section{Current experimental bounds}

\subsection{Fixed target electron  experiments}

Fixed target experiments, APEX \cite{APEX} and MAMI(Mainz Microtron) \cite{MAMI} searched for $Z'$ in electron-nucleus 
scatterings using the $Z'$ bremsstrahlung production 
$e^- Z \rightarrow e^- ZZ'$ and subsequent $Z'$ decay into electron-positron pair $Z' \rightarrow 
e^+e^- $. The absence of the resonant peak in the invariant $e^+e^-$ mass spectrum allows to obtain 
upper limits on the $Z'$ boson  coupling constants $g_{Ve}$, $g_{Ae}$ of the $Z'$ with electron. 
The A1 collaboration 
excluded the masses 
$50~MeV < M_{Z'} < 300~MeV$ \cite{MAMI} for $g-2$ muon anomaly explanation in the model with equal muon and electon couplings 
    of the $Z'$ boson with  a sensitivity to the mixing parameter up to $ \epsilon^2 = 8 \times 10^{-7}$. 
APEX collaboration used $\sim 2~GeV$ electron beam at Jefferson Laboratory and excluded masses 
$175~MeV < M_{Z'} < 250~MeV$ for $g-2$ muon anomaly explanation in the model with equal muon and electon couplings 
 of the $Z'$ boson.

\subsection{$e^+ e^-$  experiments}

BaBar experiment  has constrained   both visible and invisible $Z'$ transitions by using  $\Upsilon(1S)$ decays 
as the source of $Z'$s \cite{BaBar,BaBar0,BaBar1}.
The  search for  invisible $\Upsilon(1S)$ has been performed by reconstructing $\Upsilon(3S) \rightarrow 
\pi^+\pi^- \Upsilon(1S) $ kinematics. The bound $Br(\Upsilon(1S) \rightarrow invisible) =  (-1.6 \pm 1.4 \pm 1.6)\cdot 10^{-4}$ 
was found. Here the first uncertainty is statistical and the second systematic. In addition  
invisible  decays of $Z'$  can be searched for in radiative $\Upsilon(1S) \rightarrow \gamma + invisible $ decays.   
The decay $ \Upsilon(1S) \rightarrow \gamma + invisible$ could proceed through production of a light scalar $A$ 
followed  by its decay into invisible modes   $\Upsilon(1S) \rightarrow \gamma + A$, 
$A \rightarrow invisible$. The bound on  $Br(\Upsilon(1S) \rightarrow \gamma + invisible) $  
is  obtained at the level $(0.5 - 24) \cdot 10^{-5}$ 
assuming the  phase-space distribution for photon energy. 
Visible decays of light $Z'$ bosons were also searched for in the reaction $e^+e^- \rightarrow \gamma Z', 
~Z' \rightarrow l^+l^-(l = e, \mu)$  as resonances in the $l^+l^-$ 
spectrum. For the model with the $Z'$ boson interaction  with 
the SM electomagnetic current the mixing strength values 
$10^{-3} - 10^{-2}$ are 
excluded for $0.212~GeV < m_{Z'} < 10~GeV$.

Recently BaBar collaboration used  the reaction $e^+e^- \rightarrow Z'\mu^+ \mu^-, ~Z' \rightarrow \mu^+\mu^-$ to search 
for $Z'$ boson. The use of this process allows to restrict directly the muon coupling $g_{V\mu}$ of 
$Z'$ boson. The obtained results exclude  the $L_{\mu} - L_{\tau}$ interaction as 
possible explanation of  $(g-2)$ muon anomaly  for $m_{Z'} > 212 ~MeV$ \cite {LASTBaBar}.

The KLOE  experiment  at the DA$\Phi$NE $\Phi$-factory  in Fraskati      searched for $Z'$ in 
decays $\Phi \rightarrow \eta Z' \rightarrow \eta e^+e^-$ and   $\Phi \rightarrow \gamma Z' \rightarrow 
\gamma \mu^+\mu^-$ \cite{KLOE}. Also the reaction $e^+e^- \rightarrow Z' \gamma \rightarrow e^+e^- \gamma$ 
has been used to constrain the  $Z^`  \rightarrow invisible $ decay mode. The obtained bounds  are weaker than those from  
NA48/2 \cite{NA-48} and MAMI \cite{MAMI} bounds.

\subsection{Fixed target proton   experiments}

The NA-48/2 experiment used simultaneous $K^+$ and $K^-$ seconday beams produced by $400~GeV$ primary 
CERN SPS protons for the search for light $Z'$ boson in 
$\pi^0$ decays \cite{NA-48}.  The decays $K^{\pm} \rightarrow \pi^{\pm} \pi^{0}$ and $K^{\pm} \rightarrow \pi^0 \mu^{\pm} \nu$ 
have been used to obtain tagged $\pi^{0}$ mesons. The decays $\pi^0 \rightarrow \gamma Z^{`}$, 
$Z' \rightarrow e^+e^-$ have been  used for the search for
$Z'$ boson. $Z'$ boson manifests itself as a narrow peak in the distribution of the 
$e^+e^-$ invariant mass. 
spectrum. For the model when  the  $Z'$ boson interacts with 
the SM electomagnetic current as $L_{int, Z'} = \epsilon e Z'_{\mu}J^{\mu}_{SM}$  
the obtained bounds  exclude the $(g-2)$ muon anomaly explanation for $Z'$ boson masses 
$9 MeV < m_{Z'} < 70~MeV$ \cite{NA-48}. 

It should be noted that the decay width $\pi^0 \rightarrow \gamma Z'$ is proportional to $(g_{Vu}q_u -g_{Vd}q_d)^2 =
(2g_{Vu} + g_{Vd})^2/9$
and for the models with  nonuniversal $Z'$-boson couplings\footnote{In Ref.\cite{FENG} models with  $2g_{Vu} + g_{Vd} \approx 0 $ have 
been suggested for an explanation of recent discovery claim \cite{17mev} of   $~17~MeV$  narrow resonance  observed 
as a peak in $e^+e^-$ invariant mass distribution in nuclear transitions.}, for instance, for the model 
with $L_{\mu} - L_{\tau}$ interaction current the NA-48/2 bound \cite{NA-48} is not applicable.

\subsection{Constraints from $K \rightarrow \pi + nothing$ decay}

Light vector bosom $Z'$ can be produced in the  $K \rightarrow \pi Z'$ decay in the  analogy with the SM 
decay $K \rightarrow \pi \gamma^*$ of K-meson  into pion and virtual photon. For the model with the 
dominant $Z'$ decay into invisible modes nontrivial bound on $Z'$ boson mass and 
coupling constants arises. Namely, the BNL E949         experiment  \cite{E949}      combined with E787 results 
measured the  $K^{+}  \rightarrow \pi^{+}\nu\bar{\nu} $ decay and 
gave upper bounds on the $Br(K^+ \rightarrow \pi^+ Z')$ decay as a function of the $Z'$ mass under 
 the assumption that the $Z' \rightarrow invisible $ decay dominates. The E949 + E787 bound 
leads to the bound on the $Z'$ mass and coupling constants. For instance, in the model when $Z'$ 
couples with the SM electromagnetic current and decays invisibly modes into light 
dark matter particles,  the muon $(g-2)$ anomaly explanation due to $Z'$ is excluded for $M_{Z'} > 50~MeV $ except the 
narrow region around     $M_{Z'} =  m_{\pi} $ \cite{LM1} - \cite{LM3}. Note that in  models with 
non-electromagnetic current interactions of $Z'$, for instance when  the $Z'$ interacts with 
the $L_{\mu} - L_{\tau}$ current only, the bound from   $K \rightarrow \pi + nothing $  decay does 
not work or it could be rather weak \cite{LM2}. Recent result of the  NA64 Collaboration \cite{NA64explast} 
obtained by   using   the reaction chain $e Z \rightarrow eZZ'$,   $Z' \rightarrow invisible$    excludes the region 
$M_{Z'} \leq 100~GeV$ for muon $g-2$ anomaly explanation, see subsection 8.

\subsection{Bound from electron magnetic moment}

The experimental and theoretical values for electron magnetic moment coincide at the $10^{-12}$ level of accuracy, 
namely \cite{ge}
\begin{equation}
\Delta a_e \equiv a^{exp}_e - a^{SM}_e = -(1.05 \pm 0.82) \times 10^{-12} \,.
\end{equation}
The $Z'$ boson contributes to the $\Delta a_e$ at one loop level, see formulae (3,4). From the bound (13) 
it is possible to restrict the couplng constants $g_{Ve}$ and $g_{Ae}$. For the model 
with equal muon and electron couplings $g_{Ve} = g_{V \mu }$ and  $g_{Ae} = g_{A \mu} = 0$ 
   the $(g-2)$ muon anomaly explanation due to $Z'$ existence is excluded for 
      $M_{Z'} \geq 20 ~MeV$ \cite{gegmu}.

\subsection{Constraints from $\nu - e$ scattering}

If the  $Z'$ boson couples with the electron neutrino  and electron currents - the strongest bound arises 
from Borexino experiment  \cite{Borexino}. This experiment detects the low energy solar neutrino through the elastic scattering 
 of neutrinos on electrons \cite{Borexino}. The $Z'$ exchange modifies the SM elastic 
$\nu_e - e$  cross section 
 allowing  to obtain stringent  constraints on the $Z'$ coupling constants with the $\nu_e$ and electron \cite{Borexino.bound}. 
The obtained bound on $|g_{V\nu_e}\cdot g_{Ve}|^{1/2}$ is about $10^{-6}$ for $m_{Z'} \leq 1~MeV$ and about $10^{-4}$ for 
$m_{Z'} \leq 100~MeV$. Borexino data exclude the $(g-2)$ muon anomaly explanation in a model where  $Z'$ interacts with 
leptonic current \cite{LEE}. Also, the  data exclude the model with the $Z'$ interaction with the $B - L$ current.

\subsection{Bound from the process $\nu_{\mu}N \rightarrow \nu_{\mu} N \mu^+\mu^- $}

The neutrino trident $\nu_{\mu}N \rightarrow 
\nu_{\mu}N + \mu^+ \mu^- $ events 
allow to restrict a model where $Z'$ interacts with $L_{\mu} - L_{\tau}$ current using the results  of 
the CHARM \cite{CHARM} and the CCFR \cite{CCFR} experiments and exclude the $g -2$  
  muon anomaly explanation for $Z'$  mass $m_{Z'} \geq 400~MeV$  \cite{POSPELOV}.
 
\subsection{Beam dump experiments} 
 The results of beam dump experiments   
E137 \cite{E137}, E141 \cite{E141} at SLAC and E774 \cite{E774} at FNAL have been  used \cite{REVVEC} 
to constrain the couplings of light gauge boson $Z'$. 
In recent paper \cite{MINIboon} MiniBooNE-DM collaboration has obtained bound on  
$Y = \epsilon^2 \alpha_D (\frac{m_{\chi}}{m_{Z'}})^4 \leq 10^{-8}$.   
It should be stressed that  obtained bounds depend on unknown $\alpha_D \equiv \frac{(g_D)^2}{4\pi}$ coupling constant of the 
$Z'$ with light dark matter particles. 

The NA64  experiment \cite{NA64exp} at CERN is a fixed-target experiment  searching for dark sector particles at 
the CERN Super Proton Synchrotron(SPS) by using active beam dump technique cobined with missing-energy approach
\cite{NA64exp, NA64mu, Crivelli, Segm, Kirpich}. 
If new light boson  $Z'$ exists it could be produced in the reaction of high-energy electrons scattering off nuclei. 
The NA64 experiment uses the bremsstrahlung reaction $e Z \rightarrow e  Z Z'$
for the search for both visible, $Z' \rightarrow e^+e^-$, and invisible, $Z' \rightarrow invisible$,   decay modes. 
Also the use of the intense  muon beams for the search for $Z'$  boson in the 
 reaction $  \mu Z \rightarrow \mu Z Z' $   \cite{NA64mu} is planned in the near future. 
During summer 2016 run NA64 experiment collected approximately $2.75 \times 10^{9}$ electrons on target \cite{NA64explast}. 
Candidate events were requested to have the missing energy in the range $50 < E_{miss} <100 ~GeV$, which 
was selected based on the calculation of the energy spectrum of $Z'$ emitted in the reaction $eZ \rightarrow eZZ'$ by $e^-$  
from the EM shower generated by electron beam in the target \cite{Kirpich}. Zero events have been observed and as a consequence 
the $90\%$ $C.L.$ upper limit for the average number of signal events $N_{Z'} = 2.3$ has been derived. The obtained results   \cite{NA64explast} 
exclude the invisible $Z'$ as an explanation of the muon anomaly with masses $m_{Z'} \leq 100~MeV$ in the model with 
the $Z'$ inreraction with the SM electromagnetic current, see Fig.1. Only small mass region around $m_{Z'} = m_{\pi}$ is 
still allowed. The future NA64 runs with  $\gtrsim 10^{11}$ electrons on target can test the remaining 
mass region around $m_{Z'}  = m_{\pi}$.
%, see Fig.1.   

%\begin{figure}[tbh!]
%\begin{center}
%\includegraphics[width =0.95\textwidth]{krasnikov_nikolaifig2.ps}
%\caption{The NA64 90 \% C.L. exclusion region  in the $(m_{Z^`},~\epsilon)$ plane \cite{NA64explast}}
%\label{Fig2}
%\end{center}
%\end{figure}

\section{Model with infinite number of light $Z'$ bosons}

Models with infinite number of local fields   have been considered in Refs.\cite{1,2,0}. Note that notion of an unparticle,   
introduced by Georgi \cite{3,4} can be interpreted as a 
particular case of such models \cite{1,2,5,6,7}. So I consider a model with  infinite number of vector fields $Z'_{\mu n}$ with 
masses $m_{Z'_n}$. One can   introduce the ``effective'' vector  field $Z'_{\mu, eff} = \sum_n c_nZ'_{\mu n}$\footnote{Here $c_n$ are 
some numbers.} and  postulate the interaction of the electromagnetic field $A_{\mu}$ with the the effective field $Z'_{\mu, eff}$. 
One-loop contribution to muon  anomalous  magnetic moment is 
 $\delta a  = \sum_n |c_n^2| \frac{\alpha^`}{2\pi} F(\frac{m_{Z'_n}}{m_{\mu}}) $. The effectictive  field $Z'_{\mu, eff}$ 
represents the infinite number of vector  resonances that helps to escape bounds related with  the search for narrow resonance 
in $e^+e^-$ invariant mass distribution \cite{0}. One can speculate that  the origin of the infinite number of local vector fields 
is a  consequence  of 
compactification of some additional dimension. Namely, consider   the model with vector field $Z'_{\mu} (x,x_5)$ living in 
five-dimensional world and interacting with the four-dimensional SM  due to kinetic mixing term 
$$
L_{mix} = \frac{\epsilon}{2}  F^{\mu \nu}(x) Z'_{\mu \nu}(x, x_5 = 0) \\.
$$ 
After compactification of the $x_5$ coordinate we obtain the model (10) 
where the effective  field $Z'_{\mu, eff}$ interacts with the SM electromagnetic current.

\subsection{Conclusions}
Current  accelerator experimental data\footnote{The review of nonaccelerator bounds 
can be found in Ref.\cite{Casimir}.}  restrict rather strongly the 
explanation of the $g - 2$ muon anomaly due to existence of new light gauge boson 
but not completely eliminate it. The most popular  model where $Z'$ interacts with the SM electromagnetic current 
due to mixing $ \frac{\epsilon}{2} F_{\mu \nu}Z^{\mu\nu}$ term is excluded\footnote{
In recent paper \cite{LAST} BaBar Collaboration  used the $53~fb^{-1}$ of $e^+e^-$ collision  data 
to analyze the reaction chain $ e^+e^- \rightarrow \gamma Z^` $ , 
$ Z^` \rightarrow invisible $ and excluded completely the remaining region $m_{Z^`} \approx m_{\pi} $}.    
The Borexino data on neutrino electron elastic scattering
exclude  the models where   $Z'$ interacts with both leptonic  and $B - L$ currents.
The interaction of the $Z^{`}$ boson with $L_{e} - L_{\mu}$ 
current  is  excluded    for   $m_{Z'} \geq 214~MeV$ while still leaving the region of lower masses 
unconstrained.  More exotic models, like models with nonuniversal quark and couplings \cite{FENG}, or considered in this report model 
with infinite number of vector fields also  survive.

\section*{Acknowledgments}
I am   indebted to S.N. Gninenko and  V.A.Matveev for useful discussions and comments.

\end{document}